\documentclass[compsoc,conference,a4paper,10pt,times]{IEEEtran}
\IEEEoverridecommandlockouts
\usepackage{cite}
\usepackage{amsmath,amssymb,amsfonts}
\usepackage{algorithmic}
\usepackage{graphicx}
\usepackage{textcomp}
\usepackage{bmpsize}
\usepackage{xcolor}
\usepackage{lipsum}
\usepackage[colorlinks=true,urlcolor=black]{hyperref}
\def\BibTeX{{\rm B\kern-.05em{\sc i\kern-.025em b}\kern-.08em
    T\kern-.1667em\lower.7ex\hbox{E}\kern-.125emX}}

\usepackage{listings, xcolor}

\definecolor{verylightgray}{rgb}{.97,.97,.97}

\lstdefinelanguage{Solidity}{
    keywords=[1]{anonymous, assembly, assert, balance, break, call, callcode, case, catch, class, constant, continue, constructor, contract, debugger, default, delegatecall, delete, do, else, emit, event, experimental, export, external, false, finally, for, function, gas, if, implements, import, in, indexed, instanceof, interface, internal, is, length, library, log0, log1, log2, log3, log4, memory, modifier, new, payable, pragma, private, protected, public, pure, push, require, return, returns, revert, selfdestruct, send, solidity, storage, struct, suicide, super, switch, then, this, throw, transfer, true, try, typeof, using, value, view, while, with, addmod, ecrecover, keccak256, mulmod, ripemd160, sha256, sha3}, 
    keywordstyle=[1]\color{blue}\bfseries,
    keywords=[2]{address, bool, byte, bytes, bytes1, bytes2, bytes3, bytes4, bytes5, bytes6, bytes7, bytes8, bytes9, bytes10, bytes11, bytes12, bytes13, bytes14, bytes15, bytes16, bytes17, bytes18, bytes19, bytes20, bytes21, bytes22, bytes23, bytes24, bytes25, bytes26, bytes27, bytes28, bytes29, bytes30, bytes31, bytes32, enum, int, int8, int16, int24, int32, int40, int48, int56, int64, int72, int80, int88, int96, int104, int112, int120, int128, int136, int144, int152, int160, int168, int176, int184, int192, int200, int208, int216, int224, int232, int240, int248, int256, mapping, string, uint, uint8, uint16, uint24, uint32, uint40, uint48, uint56, uint64, uint72, uint80, uint88, uint96, uint104, uint112, uint120, uint128, uint136, uint144, uint152, uint160, uint168, uint176, uint184, uint192, uint200, uint208, uint216, uint224, uint232, uint240, uint248, uint256, var, void, ether, finney, szabo, wei, days, hours, minutes, seconds, weeks, years}, 
    keywordstyle=[2]\color{teal}\bfseries,
    keywords=[3]{block, blockhash, coinbase, difficulty, gaslimit, number, timestamp, msg, data, gas, sender, sig, value, now, tx, gasprice, origin},   
    keywordstyle=[3]\color{violet}\bfseries,
    identifierstyle=\color{black},
    sensitive=false,
    comment=[l]{//},
    morecomment=[s]{/*}{*/},
    commentstyle=\color{gray}\ttfamily,
    stringstyle=\color{red}\ttfamily,
    morestring=[b]',
    morestring=[b]"
}

\usepackage{float} 
\usepackage[linesnumbered,ruled]{algorithm2e}
\usepackage{pgf-umlsd}

\usepackage{balance}

\newcommand\project{AuthSC}

\begin{document}

\title{\project: Mind the Gap between Web and Smart Contracts}

\author{\IEEEauthorblockN{Ulrich Gallersd\"orfer and Florian Matthes}

\textit{Technical University Munich}\\
Munich, Germany \\
\{ulrich.gallersdoerfer,matthes\}@tum.de}

\maketitle

\begin{abstract}
Although almost all information about Smart Contract addresses is shared via websites, emails, or other forms of digital communication, Blockchains and distributed ledger technology are unable to establish secure bindings between websites and corresponding Smart Contracts. For a user, it is impossible to differentiate whether a website links to a legitimate Smart Contract set up by owners of a business or to an illicit contract aiming to steal users' funds. Surprisingly, current attempts to solve this issue mostly comprise of information redundancy, e.g., displaying contract addresses multiple times in varying forms of images and texts. These processes are burdensome, as the user is responsible for verifying the correctness of an address. More importantly, they do not address the core issue, as the contract itself does not contain information about its authenticity. To solve current issues for these applications and increase security, we propose a solution that facilitates publicly issued SSL/TLS-certificates of Fully-Qualified Domain Names (FQDN) to ensure the authenticity of Smart Contracts and their owners. Our approach combines on-chain identity assertion utilizing signatures from the respective certificate and off-chain authentication of the Smart Contract stored on the Blockchain. This approach allows to tackle the aforementioned issue and further enables applications such as the identification of consortia members in permissioned networks. The system is open and transparent, as the only requirement for usage is ownership of an SSL/TLS-certificate. To enable privacy-preserving authenticated Smart Contracts, we allow one-way and two-way binding between website and contract. Further, low creation and maintenance costs, a widely accepted public key infrastructure and user empowerment will drive potential adaption of Ethereum Authenticated Smart Contracts (\project).
\end{abstract}


\section{Introduction}
Users in publicly available Blockchain-based systems face a highly dangerous and hostile environment, as attackers have high incentives to steal users' funds. Although the underlying ledger provides immutable Smart Contracts, decentralized key management, and resistant consensus mechanisms, fraud, hacks, impersonation, and other malicious activities are to be encountered on a daily basis. Resolving issues of Smart Contract engineering, the formal verification of Contracts or the design of secure programming languages are subject to active research. However, an often ignored issue is the weak link to the information stored on Blockchains: The user needs to know the exact location of the Smart Contract or the externally owned account to interact with. As these addresses consist of random characters, e.g., in Ethereum 64 hexadecimal characters\cite{wood2017ethereum}, users rely on the web and a clipboard to copy and paste address information. The called address or Smart Contract contains no further information to verify whether the user interacts with the intended Contract or not.

The approach of just sharing addresses on web pages without any further connection between the business and the Smart Contract is prone to errors and can be exploited by attackers. The Coindash ICO is an example of such an attack\cite{Wieczner2017EthereumFortune}. At the height of Token Sales in 2017, Coindash conducted an Initial Coin Offering (ICO), selling their newly created token to interested investors in return for Ether (the currency of the Ethereum Blockchain). To advertise their token sale, the founders set up a website. The website contained all relevant information on how to invest their company, including the address of the Smart Contract. The sale started and many people participated; several million USD were sent in the first few hours; unfortunately, not to the intended token Smart Contract, but to a different Smart Contract set up by malicious actors, collecting all the investment funds. These malicious actors had previously hacked the website and quietly replaced the original contract address with their own, resulting in the loss of over 7 million USD. While the Wordpress content management system was compromised, the underlying web server remained intact.

To solve the issue of loose coupling between website and Smart Contract, we propose Ethereum authenticated Smart Contracts (\project). \project\ allows endorsing Smart Contracts with a private/public key pair of a TLS-certificate issued by a certificate authority (CA). This approach enables wallets or other software, which directly interact with Smart Contracts, to verify their authenticity and binding to the respective website.

\project\ contains three crucial components: the authenticated Smart Contract, which stores information about the endorsement, an off-chain verifier, which authenticates the Smart Contract as well as the \project\ registry, which prevents downgrade-attacks and offers a discovery service.

Our approach considers privacy-preserving authenticated Smart Contracts: Entities might decide to endorse Smart Contracts, however, do not want other parties to retrieve their endorsement by crawling the respective Blockchain. We expand the system to allow for an explicit discovery of such contracts: In case the user journey is not started at the website, but at the Smart Contract itself, the user is still able to authenticate this Smart Contract. This allows for varying applications, which we cover in Section \ref{applicability}. Furthermore, we discuss associated risks and potential attacks and countermeasures.

To structure our work, we organize our paper as follows: In Section \ref{architecture}, we introduce the system architecture and the methodologies to assert and authenticate Smart Contracts. We discuss privacy, risks, and attacks in Section \ref{privacy}. Section \ref{applicability} gives an overview of the applicability of the proposed artifact. We discuss the limitations and advantages of our approach in Section \ref{discussion}. We give an overview of related work in Section \ref{relatedwork} and end in Section \ref{conclusion} with a conclusion and potential future work.

\section{System Architecture}\label{architecture}
In this Section, we introduce the general approach and give an overview of the architecture, the single components, and the structure of the Smart Contracts. We further display processes of the system and design considerations. 

Our goal is to establish a binding between Smart Contracts and websites using common SSL/TLS-certificates which ensure the authenticity of web servers, the privacy, and integrity of messages between browser and server. The owner of a certificate endorses a Smart Contract by signing the respective address and storing it within the respective Smart Contract. Later, other parties can retrieve the certificate from the Blockchain and validate it against their trusted certificate authority servers, thereby authenticating the contract. Afterward, they can engage with the specific Smart Contract or further evaluate data stored in the Smart Contract, aware of the actual identity of the counterparty. A reference interface is available online in\cite{AuthorCitation2019AuthorCitation}, using Solidity for Ethereum. In theory, \project\ is Blockchain-independent; any other network can profit from this methodology.

To enable identity assertion and authentication, our system consists of three components: 
\begin{itemize}
\item The \textbf{On-Chain Authenticated Smart Contract} contains methods for storing and updating signature information. This functionality is provided as a library and therefore independent from the actual Smart Contract (e.g., token contract or other purposes).
\item The \textbf{Off-Chain Verifier} is an application that runs outside of the Blockchain. It is responsible for verifying on-chain assertions. It retrieves data from the certificate issuer, the website owner, the Blockchain, and potential certificate authorities. With this information, it enables authentication of on-chain information in the context of the user-defined environment. 
\item The \textbf{Authenticated Smart Contract Registry} prevents downgrade-attacks by providing a list of Smart Contracts for domains. Further, it optionally allows for easy discovery of Smart Contracts which implement this library. It manages a list of existing Smart Contracts and their respective domains such that interested parties can find these Smart Contracts. 
\end{itemize}

\begin{figure*}
  \includegraphics[width=\linewidth]{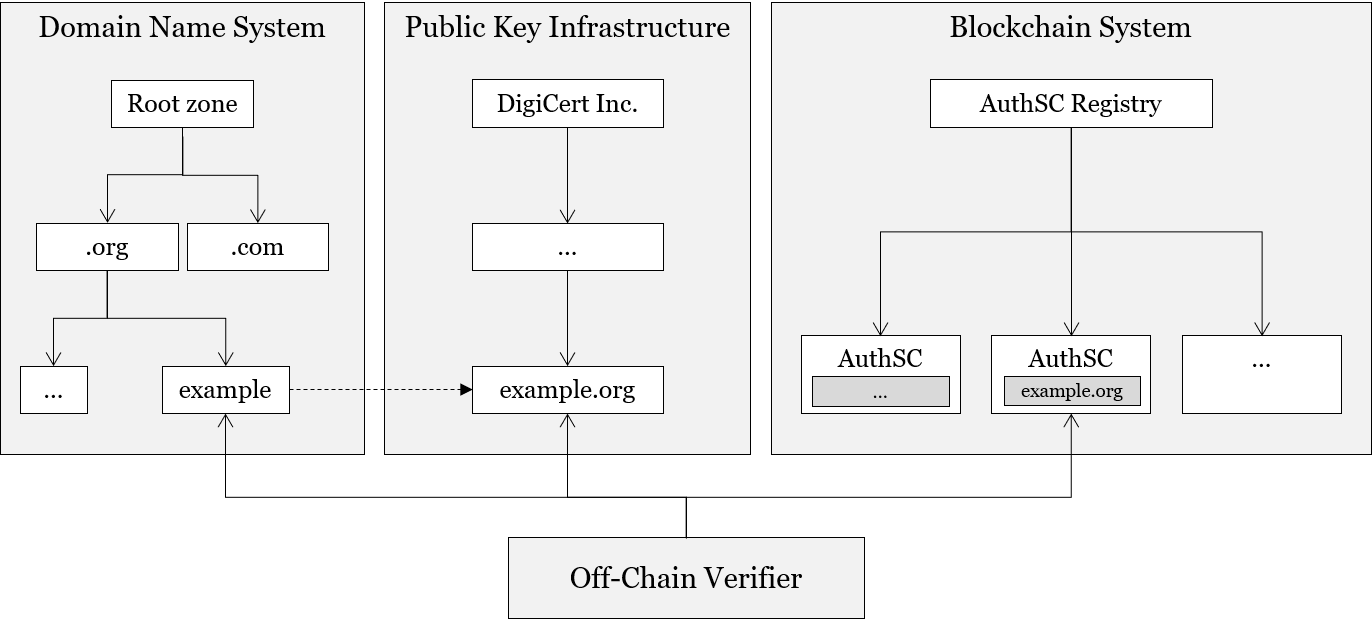}
  \caption{High-Level Structure.\label{highlevel}}
\end{figure*}

Figure \ref{highlevel} depicts the overall structure of the proposed architecture. The On-Chain Authenticated Smart Contract (\project) is placed within the Blockchain system and referred to by the Authenticated Smart Contract Registry. The Off-Chain Verifier accesses the \project, the DNS, and the PKI to verify the correctness of all obtained data and to authenticate the specific Smart Contract.

In the remainder of this Section, we analyze the responsibilities and methodologies of the single components.

\subsection{On-Chain Authenticated Smart Contract}

An entity asserts its identity to a Smart Contract by binding it to a public domain. To allow such identity assertion, the contract has to store the following data for authentication: 
\begin{itemize}
    \item the fully qualified domain name for later identification (optional)
    \item the signature information (the address of the Smart Contract signed by the private key)
\end{itemize}

Additional information required for the authentication of the Smart Contract is retrieved from external data sources. This identity assertion functionality can be embedded in an arbitrary Smart Contract, allowing \project\ to be used in any on-chain environment. The interface a Smart Contract has to implement to adhere to \project\ is depicted in Listing \ref{lst:code}. Please note that an FQDN is provided as an array, e.g., \textit{hq.example.org} translates to \texttt{["org", "example", "hq"]}. This approach allows for easy access to parts of the FQDN.

\begin{lstlisting}[language=Solidity, caption=On-Chain Smart Contract Interface in Solidity, label=lst:code]
interface ERCXXX /* is ERC165 */{

    // @dev Event is emitted when the FQDN changes
    event FQDNChanged(string[] indexed FQDN);
    
    // @dev Event is emitted when the signature changes
    event SignatureChanged(string signature);

    // @dev Returns the domain for the given SC
    function getFQDN() external view returns (string[] memory);
    
    // @dev Returns the signature for the given SC
    function getSignature() external view returns (string memory);
    
    // @dev Allows to set the FQDN
    function setFQDN(string[] calldata _FQDN) external;

    // @notice Sets the signature (containing the SC address)
    function setSignature(string calldata _signature) external;
}
\end{lstlisting}

Three steps are required to enable authentication of the Smart Contract: Smart Contract creation, signature generation, and signature upload. For simplicity, we omit intermediate or supporting activities such as creating wallet addresses, funding the accounts, and including the library in the respective Smart Contract. 

\begin{enumerate}
\item \textbf{Smart Contract Creation:} As it is impossible to update previously existing Smart Contracts, one has to create a new Smart Contract upfront. A contract has to be instantiated with information on the address of the owner of the Smart Contract and optionally the domain to which the owner has access to (e.g., the domain or any subdomain). If a domain is not provided, two-way binding is not supported, which we discuss in Section \ref{privacy}.

After the first step, the Smart Contract is set up on the Blockchain.

\item \textbf{Signature Generation:} The signature is still missing in the Smart Contract and thus, the SSL/TLS certificate has not endorsed this Smart Contract, as any party can execute the first step for any domain. The entity retrieves the unique contract address and signs this string with its private key of the SSL/TLS-certificate. As the signature contains only one specific Smart Contract address, other Smart Contracts cannot use this information to pretend to be also endorsed by this domain.
\item \textbf{Signature Upload:} The signature created in the previous step is transmitted to the Smart Contract as part of a regular transaction calling the method \texttt{setSignature}. The contract validates that the transaction is indeed created by the original owner of the Smart Contract, as the contract itself is not able to verify if the provided data is correct. If the owner issued the transaction, the payload is stored in the respective field of the Smart Contract, awaiting later retrieval.
\end{enumerate}

It is possible to reduce the process to two steps, as it is possible to sign the address of the contract before its initialization, providing the signature alongside the domain name. This is possible as the addresses of to-be generated Smart Contracts are deterministic. For simplicity and to remove possible interference with other transactions issued from the same address, we decided to first upload the contract and then actually submit the signature data, preventing that an incorrect address is signed.

\subsection{Off-Chain Verifier}

Smart Contract computations must be deterministic. Thus, a Smart Contract cannot request external data like lists of CAs or web servers, as a recurring calculation (e.g., in another full node in the network) might lead to a different result. Therefore, we move the authentication process of the endorsement stored within a Smart Contract off-chain. The software itself needs to access the following data sources:
\begin{itemize}
\item The address of the relevant Smart Contract, which is obtained via the web or is obtained by the Authenticated Smart Contract Registry or proprietary discovery services.
\item The contents of this Smart Contract, e.g., optionally the domain name and the signature
\item The signed certificate, its public key alongside the information of the certificate authority, obtained by requesting the domain and retrieving the SSL/TLS-certificate
\item The list of trusted CAs of the user
\end{itemize}

It is possible to store the entire SSL/TLS-certificate in the Blockchain, rendering the request to the web server unnecessary. However, as these certificates take additional space, we oppose to storing them on-chain for cost reasons. If decentralized storage systems such as the Interplanetary File System (IPFS)\cite{Benet2014Ipfs-contentSystem} gain practical relevance, it becomes cost-efficient to store complete certificates in decentralized systems, no longer relying on web servers of the respective owners. As long as such systems remain in an early stage, we continue to request the certificate from the respective servers.  

The list of trusted certificate authorities is defined by the user, either directly by providing a list themselves or by the certificate authority list stored in her/his computer or browser. 

Smart Contract authentication involves four steps. Again, we omit intermediate steps and assume a fully accessible system, also we assume that the contract address is known. 

\begin{enumerate}
\item \textbf{Smart Contract Information Retrieval:} The application first retrieves all relevant information for later authentication. It collects the domain and signature data from the Smart Contract. 

\item \textbf{Certificate Retrieval:} Afterwards, it connects to the respective domain given in the Smart Contract (or previously known by the website) and obtains the certificate.

\item \textbf{Smart Contract Signature Verification:} The software validates if the Smart Contract address is signed by the private key of the certificate. As the certificate has to belong to the website, no additional check is required. If the signature is correct, the software proceeds with step 4, otherwise it aborts the identity authentication with an error.

\item \textbf{Certificate trust:} After the successful verification of the signature, the software checks whether a trusted certificate authority has (indirectly) signed or issued the certificate found in the Smart Contract and the web server. If the certificate and its public key is signed by a trusted CA, then the identity is successfully validated, otherwise, the program aborts with an error. It executes the certification path validation algorithm as defined in \textit{RFC 5280}\cite{Cooper2008RFCProfile}.
\end{enumerate}

\begin{figure*}
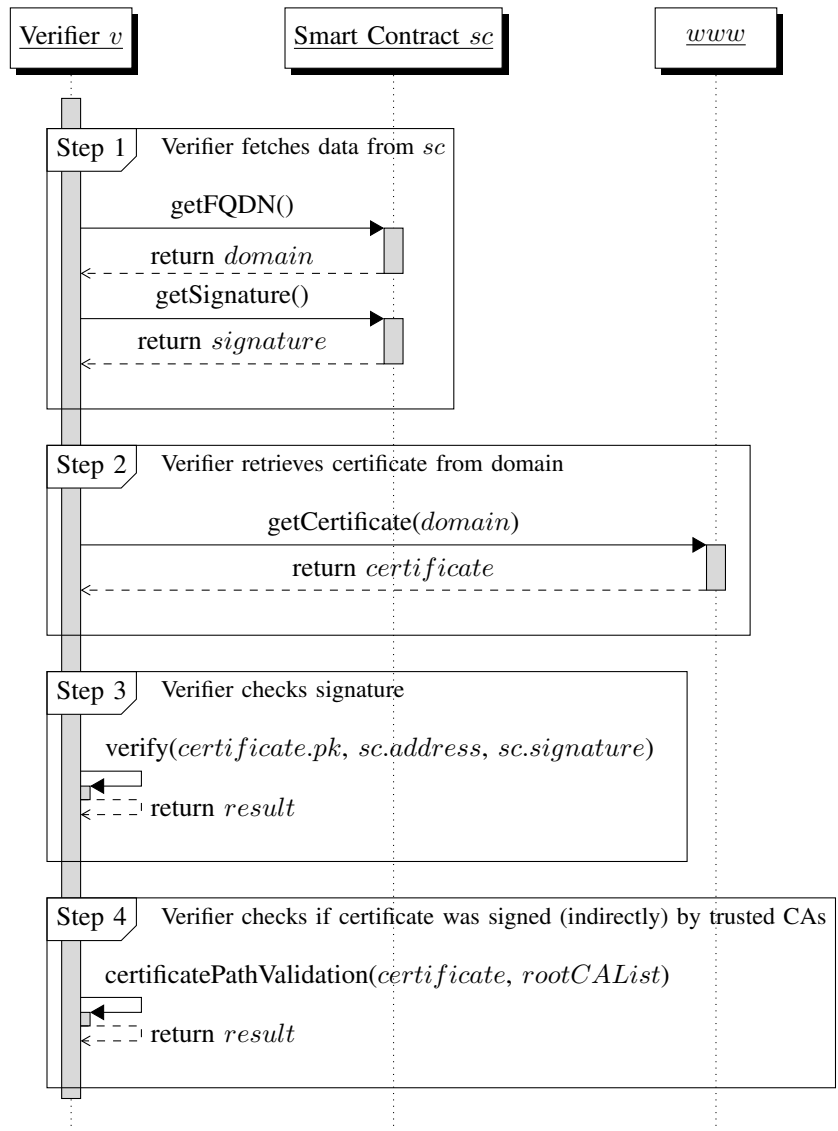

  \centering
  \begin{sequencediagram}
    \newthread{A}{Verifier $v$}{}
    \newinst[2]{B}{Smart Contract $sc$}
    \newinst[2]{C}{$www$}
    
    \begin{sdblock}{Step 1}{Verifier fetches data from $sc$}
    \begin{call}{A}{getFQDN()}{B}{return $domain$}
    \end{call}
    
    \begin{call}{A}{getSignature()}{B}{return $signature$}
    \end{call}
    \end{sdblock}
    
    \begin{sdblock}{Step 2}{Verifier retrieves certificate from domain}
    \begin{call}{A}{getCertificate($domain$)}{C}{return $certificate$}
    \end{call}
    \end{sdblock}
    
    \begin{sdblock}{Step 3}{Verifier checks signature}
    \begin{call}{A}{verify($certificate.pk$, $sc.address$, $sc.signature$)}{A}{ return $result$}\end{call}
    \end{sdblock}

    \begin{sdblock}{Step 4}{Verifier checks if certificate was signed (indirectly) by trusted CAs}
    \begin{call}{A}{certificatePathValidation($certificate$, $rootCAList$)}{A}{ return $result$}\end{call}
    \end{sdblock}

  \end{sequencediagram}
  \caption{Authentication of a Smart Contract attributed identity}\label{authenticationdiagram}
\end{figure*}

Figure \ref{authenticationdiagram} depicts the steps involved in the authentication process in detail. The definitions for digital signature methods are well established\cite{AuthorCitation2019AuthorCitation}. For this algorithm, we assume that the Smart Contract address is previously known.

\subsection{Authenticated Smart Contract Registry}

The previously described on-chain architecture is comprised of a single type of Smart Contract. A set of such Smart Contracts is sufficient to create and manage identities. However, it is relevant to know 1) whether a contract exists for a given domain and 2) under which address it is deployed. To find such contracts or check if contracts exist, we propose extending this architecture by introducing an Authenticated Smart Contract Registry, which allows to quickly lookup all domains registered on the Blockchain. This registry lists all contracts which adhere to this interface standard and claim to be the identity contract for a specific domain. We need to modify such a registry for two reasons: 1) Our on-chain data does not converge to a single truth, but allows creating different perceptions of reality. 2) A company can decide to offer multiple Smart Contracts, providing various services, with their identity tied to them. We, thus, allow that multiple Authenticated Smart Contracts can exist for the same domain. Even though it is possible to search the complete Blockchain for such contracts\cite{AuthorCitation2019AuthorCitation}\cite{Frowis2019DetectingEthereum}, it is easier and faster to use an on-chain Smart Contract. 

The registry Smart Contract is in place to map domains to Smart Contract addresses by a) storing SHA3-hashes of domains with respective Smart Contracts and b) domains with respective Smart Contracts for reasons we discuss in Section \ref{privacy}. The usage of such contract is as follows.

\begin{enumerate}
\item \textbf{Insertion of Contract Information:} All parties controlling authenticated Smart Contracts submit information about their contract and optionally the respective domain to the discovery service Smart Contract. 
\item \textbf{Domain Lookup:} If a user searches for a domain, s/he queries the registry Smart Contract and asks for all contracts which are assigned to that specific domain or the hash of that domain. The contract returns all relevant contract addresses.
\item \textbf{Contract authentication:} The client can execute the previously described authentication method for each of the returned contracts, ensuring that the correct Smart Contract, if it exists, is found. 
\end{enumerate}

This Registry Smart Contract relies on owners submitting their created contracts. Thus, 1) ideally every authenticated contract is added and 2) the amount of incorrect data is minimized. First, the entities creating contracts have a strong incentive to be found, as they want users and other parties to interact with their Smart Contract, otherwise they would not have deployed it on the Blockchain. The discovery service Smart Contract allows them to advertise their service and enable users to find them. Second, malicious entities which do not own the respective TLS/SSL-certificate are discouraged from linking respective contracts as it costs money to add the information, the authentication of these contracts will fail and the users will not even notice the existence of such contracts, as modern computation power is sufficient to process hundreds of such contracts in milliseconds. Given this (dis)incentive structure, we allow every entity in the network to add data to the Authenticated Smart Contract Registry.

\section{Privacy Concerns, Risks and Attacks}\label{privacy}
In this Section we discuss privacy considerations, one-way and two-way bindings, key management risks and downgrade-attacks on \project.

\subsection{Privacy Considerations}
The design goal of open and transparent Blockchains are inherently contrary to the goal of privacy of user activity in such networks, as all transactions, messages and deployed code is available to every node in the system. We are aware that it is possible to crawl all Smart Contracts stored in a Blockchain and also identify Authenticated Smart Contracts. We acknowledge that some parties might want to use the functionality provided by \project, but do not want to publicly state that this is the case. For that, we introduce one-way and two-way binding between FQDN and Smart Contract.

\begin{itemize}
\item One-way binding means that only the website links to the Smart Contract; the contract itself does not link to the website. Without knowledge about the domain, the Smart Contract itself cannot be authenticated, because it is unclear against which domain it should be authenticated. This allows for a privacy-preserving authenticated Smart Contract, resulting in a contract that is findable on-chain, however, cannot be attributed to a domain (and therefore an owner). To enable one-way binding, the Smart Contract itself does not store the domain information. Nonetheless, to prevent downgrade-attacks, it is recommended to store the address and the hash of the domain within the registry Smart Contract as outlined in Section \ref{downgrade}.
\item Two-way binding results in Smart Contracts which can be authenticated by only knowing the respective Smart Contract address. In this case, the Smart Contract stores the FQDN within the Smart Contract, allowing to resolve the certificate which signed the Smart Contract address. In this case, it is also advised to store the hash of the domain (to prevent downgrade-attacks) as well as the domain itself (to enable discovery) within the Authenticated Smart Contract registry.
\end{itemize} 

\subsection{Key Management Risks}
The key material in TLS/SSL-certificates faces the following risks:
\begin{itemize}
    \item \textbf{Key Material Expiration:} Key material in X.509 certificates expires regularly. Public key and signature information have to be updated, otherwise, the verifying entity cannot assert the correctness of the authenticated contract. To update the Smart Contract, steps 2 and 3 of the assertion process have to be executed again with the new certificate. This procedure ensures that the previously submitted information stored in the contract remains valid. If an entity wants to invalidate her/his contract, s/he updates the respective SSL/TLS-certificate, thus invalidating the signature stored in the contract.
    \item \textbf{Certificate Key Revocation:} Key revocation becomes mandatory if key material gets compromised. Such key material could be used to create authenticated Smart Contracts, potentially tricking users into believing that the new Smart Contract does indeed belong to the compromised entity. However, as our approach relies on the public key of the web server, an replacement of the keys stored on the server will also render potential deceiving Smart Contracts invalid. The software is not able to verify Smart Contracts with old keys. Further, the software could also evaluate whether the to-be verified certificates are included in so-called certificate revocation lists (CRL). This becomes relevant when the complete credential is stored on-chain, as the additional check with the web server is omitted. 
\end{itemize}

\subsection{Downgrade-Attacks}\label{downgrade}
We found that \project\ might be vulnerable to downgrade-attacks.

In a downgrade-attack, an adversary tricks a party in a communication protocol to assume that the other party is incapable of adhering to newer (and more secure) versions of the communication protocol, leading to the usage of an older (and less secure) version of the protocol. As an older version is susceptible to further attacks or has no protection at all (e.g., plain text), an adversary can further exploit the communication. The same is also true for \project. Looking at the base case and our introductory example of CoinDash, it becomes apparent that the user needs to know that a website uses \project\ to protect their users from sending transactions to malicious contracts. However, we cannot expect a user to know whether the counterparty actually implements \project\ and in case of CoinDash, an address swap is still possible without any further mechanisms in place.

Therefore, the Off-Chain Verifier (or other software implementing the authentication mechanism) needs to know whether a contract exists for a given domain. If the verifier obtains a contract-address from a website, it checks if this contract supports the \project\ interface and, if that is not the case, another Smart Contract exists for that domain and further if that Smart Contract can be verified. This verification is necessary to prevent Denial of Service attacks, as otherwise attackers could register Smart Contracts for any domain and harm the communication between the website and their users. If such a contract is encountered without the current Smart Contract adhering to the interface standard, a warning is emitted to the user, stating that the current contract has no additional protection against impersonation and that other contracts with such protection exist.

The Authenticated Smart Contract Registry is asked about already existing Smart Contracts for a domain. If the original and authenticated Smart Contract is registered within this registry, downgrade-attacks can be prevented. 


\section{Applicability}\label{applicability}
In this Section, we briefly introduce potential usages for \project, allowing further usages for authenticated Smart Contracts.

\subsection{Consortia Member Identification}
Companies and other entities often opt for permissioned settings concerning Blockchain systems: Either they preselect nodes that are allowed to produce blocks in the network (public permissioned Blockchains) or they limit the accounts which are allowed to interact with a given Smart Contract architecture\cite{deKruijff2017UnderstandingOntologyb}. Often, these entities (Smart Contracts or externally owned accounts) are white-listed after an off-chain proof; usually, teams from different companies communicate their account information outside of the Blockchain network. However, this process is expensive, as for every new connection trust has to be established. To enhance this process, proofs can be generated with \project\ that endorse accounts and Smart Contracts, allowing the "administrative" party to automatically add new entities to the network after sufficient proof. This not only reduces costs but allows for novel filtering, e.g., any domain with a TLD from e.g., Sweden is allowed to participate in the network. Proofs can be stored in public permissionless Blockchains (e.g., the Ethereum Mainnet) and used in private Blockchains, allowing an even further adoption of SSL/TLS-certificates within these networks. 

\subsection{Information attribution and authentication in public Blockchains}
The properties of immutability, transparency, and longevity of data and information are often used as arguments for Blockchain technology, as many companies value such properties for their use cases, e.g. reduce fraud or increase transparency towards their users. Data stored on Blockchains with the intention of transparency, however, lack the attribution or the connection to the company, such that a proof that data comes from a company or institution is bothersome, as prior communication is required as evidence. For these companies, it is much easier to authenticate this Smart Contract, store their data in it, and let other parties validate the stored data alongside authenticating the Smart Contract. To give an example, this approach could be used for allowing online-authentication of digital certificates, e.g. degrees or references. Hashes of such documents are stored on the Blockchain, similar to other Blockchain-based approaches\cite{Schmidt2016BlockcertsAnBlockchain}. The authenticated Smart Contract ensures the correctness of the identity of the issuer. Third parties (e.g., potential employers) can compare the hashes of received documents to the hashes contained in the Smart Contract. This software additionally authenticates the Smart Contract and validates the asserted identity. With such a solution in place, software can automatically scan if uploaded documents by the applicant are valid or not.

\section{Discussion}\label{discussion}

In this Section, we discuss the limitations and advantages of our approach to better understand the implications when using this artifact.

\subsection{Limitations}\label{limitations}

To start with, this approach only supports the authentication of domain owners with certificates, e.g., businesses or users with own websites, but not private persons or any other parties without such certificates. This is a caveat, as other parties, e.g., users, are also interested in having authenticated addresses or Smart Contracts. Sending funds to other parties or friends should involve some form of authentication, unfortunately, our approach is not suitable for this. Solving the issue for regular users would also result in reduced onboarding costs if the authenticity of the user is required, e.g., in financial transactions or products. However, identity providers could use \project\ to authenticate their contracts to later assert identities to their users, such that they can use authenticated addresses later on with other service providers.

The \project\ approach relies on SSL/TLS-certificates issued by certificate authorities. The utilization of such certificates implies that these institutions are trustworthy and that limitations of this hierarchical public-key infrastructure also apply to \project\ regards to centralization. However, the transparent storage of signatures of Smart Contract addresses can lead to increased trust in the infrastructure, as outlined in Section \ref{conclusion}. Furthermore, as the basis of information is the web (which is deemed to be trustworthy), we provide similar security.

Blockchain and Smart Contract technology is a growing area of business and research, and a lot of different solutions compete for users, companies, and funds. At the current stage of development, it is impossible to predict which authentication or identity solution will prevail in the long run and even if such technologies will have relevance in the future at all, and this also applies for \project. However, if the usage of Blockchain technology increases, such solutions will be required or even enable such usage increase. 

The system is designed such that different entities can reach different conclusions about the validity of on-chain data. Consequently, users verify data “off-chain”, not resulting in “absolute truth” on-chain. It is possible to have different perceptions of reality. Further, an entity may not want to engage with another entity, even if it knows it’s true identity.

Our presented approach requires rewriting existing contracts to adhere to \project's interface, potentially imposing barriers to companies. Furthermore, access to the SSL/TLS-certificate is required. Every company that uses the \project-approach has to follow the setup process, incurring costs up-front. As many businesses rely on web hosting companies, access might not be trivial, as leakage of the certificate must be prevented. 

\subsection{Advantages}

Our system uses, in contrast to all other approaches outlined in Section \ref{relatedwork}, existing signature, and certificate authority schemes. This has a two-fold benefit: First, there is a straightforward onboarding process in place, as existing structures are used. Entities do not have to go through additional Know-Your-Customer-processes (KYC) to authenticate their Smart Contract, it is inherently existent if they own a domain with an SSL/TLS-certificate. No entity is the gatekeeper of this approach and any entity worldwide can immediately use this approach, enabling a completely open system. Second, the system itself does not need a bootstrapping process or the persuasion of other entities to join and create the desired network-effect. 

SSL/TLS-certificates are a passive form of authentication. A user requests a website and this website returns the data with the respective certificate. The web server cannot use its certificate in communication as the initiating party, e.g., contacting another web server and being authenticated by the TLS/SSL certificate. With our approach, the SSL/TLS-certificates can be used actively, as other Smart Contracts can allow for transactions from such authenticated contracts, ensuring the authenticity of their counterparty. This is a novelty for world-wide-web based certificates.

Our system enables direct commercial relationships between two businesses, worldwide, as until now there is no way for companies to engage directly over Distributed Ledger Technologies. Both parties can authenticate themselves with their public key and signature in their respective Smart Contract and signal the other party the willingness to engage in further trade activities. Additionally, any other party owning a domain and an SSL/TLS-certificate such as governments, NGOs, or private persons can participate in this scheme, allowing for an open ecosystem. 

In some existing identity solutions, a central entity defines the correctness of identities and their claims. This results in a single point of failure. If the central entity fails, the entire system is rendered useless or even worse, is open to manipulating arbitrary data. In our approach, the users can, if they want, decide on their own which entities to trust, resulting in their perception of what information is correct or not. Other users who are not familiar with the system or technology can use the standard configuration, resulting in a valid perception. 

If a company using our approach implements a use case familiar to the credential use case, the experience of potential users visiting the website is similar to other non-Blockchain websites. For the identification and the authentication, no additional tools like MetaMask or other wallet software are required, as every used technology is already available in modern browsers. Therefore, the complexity of the Blockchain and the implemented system is abstracted from the user; if s/he is not interested in the functionality of the system, chances are high that s/he will not notice that the system is Blockchain-based.

The overall costs of our approach are low. Excluding the costs of requesting a certificate, a company (or an entity creating an authenticated contract) faces three primary cost drivers:

\begin{itemize}
\item The deployment of the Smart Contract on the Blockchain (required only once)
\item The insertion/update of certificate material (only required if the certificate changes)
\item The usage of the actual intended Smart Contract functionality (not covered in our calculation, as these costs apply regardless of the authentication)
\end{itemize}

As the number of computational steps of each of the first two processes depends on several parameters (e.g., domain length, enabled, or disabled compiler optimization), costs cannot be determined precisely upfront. To give a range, deploying an exemplary contract\cite{AuthorCitation2019AuthorCitation} uses about 900,000 to 1 million Gas units. Currently (233 USD / Ether, 1st March 2020), ETHgasstation.info recommends a Gas price of 8 GWei for fastest and 1 GWei as a safe minimum for executing transactions. Selecting the minimum price of 1 GWei as Gas price, the costs amount to approximately 0.23 USD, a high price of 6 GWei amounts to 1.86 USD for the first step. In the second step, the insertion of key material takes about 10\% of the amount of Gas of the contract creation. In total, an investment of less than 5 USD is sufficient to get started with this approach, enabling a broad audience to use this system. The detailed prices are depicted in Table 1.

\begin{table}[H]
\caption{Costs of Smart Contract Interaction}
  \begin{tabular}{@{}|l|c|c|@{}}
    \hline
        & \textbf{Deployment} & \textbf{Key Material Insertion} \\ \hline
    Gas usage & 1,000,000 Gas & 120,000 Gas \\ \hline
    Safe Low (1 GWei) & 0.233 USD & 0.028 USD \\ \hline
    High Speed (8 GWei) & 1.864 USD & 0.224 USD \\ \hline
  \end{tabular}
\end{table}

\section{Related Work}\label{relatedwork}
In this Section, we outline concepts that partly overlap with our research or try to solve similar issues. As our approach lies within multiple domains, we categorize related work in three distinct categories: 1) verification and validation of Ethereum addresses, 2) name services, and 3) decentralized identity or identity attribution systems.

\subsection{Verification of Ethereum addresses}
We focus on the verification of Ethereum addresses, however, these approaches remain identical for addresses of other Blockchains or public-key address schemes.
\begin{itemize}
\item Address Checksums: In EIP55, an address scheme was introduced which capitalizes parts of the characters in the address to allow for an easy verifyable checksum\cite{Buterin2016Mixed-caseEncoding}. It helps to prevent mistakes (e.g., left out characters) while copying the addresses. It is often used inside wallets (e.g., MetaMask). However, as an attacker can easily generate such checksums as well or the tools used to generate them automatically, the checksum does not help the user to distinguish between valid and illicit Smart Contracts.
\item Vanity addresses: Vanity addresses are specially crafted addresses to enable a partly readability. For example, the author of Profanity, a tool for producing these vanity addresses, has given 0x\textbf{000dead000}ae1c8e8ac27103e4ff65f42a4e9203 as donation address\cite{johguseProfanityEthereum}. The software computes private keys with a respective public key until an address has been found which satisfies the predefined requirements. Alternative approaches include the usage of account nonces (specific to Ethereum) to increase speed, however, the result remains identical. An attacker is also able to generate such an address with similar patterns; given s/he has the computing power.
\item Account address images: An approach used by some wallets is blockies. Blockies is a library that allows generating unique images of addresses, called identicons\cite{dachtlerBlockies}. These images should help users to identify single addresses and better recognize if addresses have been swapped. However, also this approach suffers from potential brute force to recreate almost identical images, as work by Austin Griffith shows \cite{VanityMedium}.  
\item Multitude of address displayed: Some popular tokens, such as Token-as-a-Service (TaaS), approached the problem by introducing redundancy by displaying the address in multiple settings (website, mail, and others) and formats (image, text, and others)\cite{Taas}. This approach increases resilience against attacks, as attackers might not able to change addresses in already sent emails. However, it shifts the responsibility to the user. S/he is responsible for the tiresome comparison and verifying the correctness of the address. If the user does not compare these addresses, an attack could still be successful.
\end{itemize}

\subsection{Name Services}
Name services are systems that allow attributing human-readable names to complex and changing information, e.g., IP-addresses. This is a well-known approach and is also the basis for today's Domain name system \cite{Mockapetris1988DevelopmentSystem}, as it maps (amongst other parameters) between FQDN and IP-addresses.

\begin{itemize}
\item Ethereum Name Service (ENS): ENS is a well-known service that allows the registration of domains for the TLD .eth\cite{ens}. It started in 2018 and currently, over 700,000 domain auctions have happened. Users bid for domains in a sealed auction and after a predefined time, the ownership is transferred to the bidder with the highest offer. The funds are locked for the time the domain is registered. It can be refunded once the ownership is abandoned. The approach is promising and finds a lot of users within the Ethereum community, however, the overall adoption is practically non-existent. There is no judiciary system that allows reassigning domain names, e.g., in cases of trademark issues. Likely, popular domains registered in ENS do not link to respective company accounts. Further, the "root zone" is managed by a multi-signature wallet owned by seven single individuals.
\item Other PKI-implementations: There is a series of approaches trying to replicate public key infrastructures on Blockchains, creating such systems from scratch. The key difference between these approaches and ours is that we include preexisting infrastructure in our approach, eliminating the need for any bootstrapping. Out of brevity, we do not cover single approaches but rather link to already existing research conducted in \cite{brunnercomparison}, which provides an overview of several implementations and their key differences. 
\end{itemize}

\subsection{Decentralized Identity and Identity Attribution}
Active research is conducted in the area of decentralized identity, web of trust, and verifiable credentials. W3C's emerging standard on Decentralized Identifiers (DIDs) and Verifiable Credentials (VC) receive a lot of attention \cite{CredentialsCommunityGroup2019DataDIDs}. The basic idea is to enable credentials independent from any third party, e.g. a website or a provider (e.g. \textit{Login with Google}) by storing a DID on a distributed ledger. These ledgers (e.g. Ethereum) need to provide methods to resolve such DIDs to DID documents which have additional information about cryptographic mechanisms, keys, and verification endpoints. The research focuses on standardizing interfaces required for interacting with varying verification endpoints. These systems primarily focus on enabling personal identity without the requirement for a third party. In these systems, Blockchain functions as a distributed immutable storage; it is not intended to assign identities to Smart Contracts or on-chain identities. Furthermore, DIDs lack widespread adoption, and bootstrapping is required.

\section{Conclusion and Future Work}\label{conclusion}

In this paper, we introduce a novel approach to use established forms of certificate authorities and signature schemes to allow for a secure binding between Smart Contract-based Systems and the web. In comparison to other approaches, our method does not depend on the adoption rate or a network effect of the solution, as SSL/TLS-certificates and the cryptographic mechanisms are omnipresent and heavily used in today’s world wide web. The low costs and simplicity for the end-user make the usage of SSL/TLS-certificates in the context of Blockchain reasonable.

We identify three major directions of future work: risk mitigation of using TLS/SSL-certificates, integration of \project\ into already existing identity management systems, and enabling on-chain identity.

As discussed in Section \ref{limitations}, \project\ inherits already existing drawbacks of the public key infrastructure, such as revoked certificates, reissued certificates (by malicious actors), or spoofing attacks with stolen keys. To resolve such issues, they need to be addressed in the underlying PKI. However, the increased transparency in the storage can support algorithms in detecting abnormal issuance of certificates or re-issuance of already trusted Smart Contracts. Analog to attempts like Certificate Transparency \cite{Laurie2014CertificateTransparency.}, transparency of \project\ Smart Contracts (and thus signatures) could enable heuristics to mitigate risks inherited from the underlying layer.

The integration of this approach into already existing identity management systems like the upcoming Decentralized Identifiers (DID) W3C standard\cite{Reed2019DecentralizedIdentifiers} can be promising. A DID-method specifically tailored to account for SSL/TLS-certificates could help to reboot the Web of Trust, as it would provide potential trust anchors without the requirement for bootstrapping.

The on-chain authentication of identities should be a significant direction of future research, as it enables the full potential of this approach, allowing any use case which involves on-chain computation to be implemented and backed by this approach, enabling the usage of identity on-chain. It merely allows for binding between the web and Smart Contracts, but further integrates TLS/SSL-certificates as an identity provider into Smart Contract systems. From our perspective, there are three potentials of how such an approach can be facilitated: 

\begin{enumerate}
\item Different types of certificates: The potential forms of SSL/TLS-certificates are varying: EV-certificates, wildcard certificates, or even multiple certificates for a single domain require additional analysis of the current certificate ecosystem to enable the usage within \project\cite{Levillain2012OneMeasurement}. The usage of TLS proxies especially present in enterprises requires further consideration\cite{Waked2018TheAppliances}, further invalid, and malformed certificates and their impact on security needs further consideration\cite{chung2016}.
\item Enabling self-hosted oracles for validation of certificates: Each owner of a Smart Contract who wants to identify other Smart Contracts defines her/his oracle which is responsible for validating certificates s/he wants to engage with. By self-hosting such oracle, the entity can tremendously reduce the costs in comparison to a full on-chain authentication and define its own rules and trusted CAs. This is a simple solution; however, it induces the necessity of off-chain oracles.  
\item Migration of existing CA structures to Blockchain-based systems, such that all data is stored safely and in a transparent manner on the chain. If existing on-chain key mechanisms are in place, such an approach could reduce the costs of true on-chain authentication to a minimum, enabling a truly trustless and widespread usage worldwide. 


\end{enumerate}
\bibliographystyle{unsrt}
\balance
\bibliography{submission}
\end{document}